\documentclass[12pt]{article}
\usepackage{amsmath}
\usepackage{amssymb}
\begin{document}
\vskip 2cm
\begin{center}
{\bf {\Large Quantum Symmetries  and Conserved Charges  of the Cosmological Friedmann-Robertson-Walker Model }}\\

\vskip 3.2cm

{\sf B. Chauhan}\\\vskip 0.4 cm

{\it  Department of Physics,  Centre of Advance Studies, Institute of Science,}\\ \vskip 0.1 cm
{\it Banaras Hindu University, Varanasi - 221 005, (U.P.), India}\\

\vskip 0.3cm


\vskip 0.1cm

{\small {\sf {E-mail: bchauhan501@gmail.com}}}

\end{center}

\vskip 3cm

\noindent
{\bf Abstract:}  
We discuss both the off-shell nilpotent Becchi-Rouet-Stora-Tyutin (BRST) and anti-BRST symmetry transformations
 for the cosmological Friedmann-Robertson-Walker (FRW) model with a differential gauge condition
 in the extended phase space. In this discussion, the presence of anti-BRST symmetry provides the
 complete geometrical description of BRST within the ambit of the supervariable approach. We derive the conserved 
(anti-)BRST charges for the FRW model using the celebrated Noether theorem and show the nilpotency 
and absolute anti-commutativity properties of these conserved charges within the realm of BRST formalism.
 Finally, we prove the sanctity of (anti-) BRST symmetries through the derivation of these symmetry
 transformations within the framework of the (anti-)chiral supervariable approach (ACSA) to BRST formalism.    \\

\vskip 1.2cm



\newpage
\noindent
\section{Introduction}

\vskip 0.4 cm

The gauge theories provide theoretical description of {\it three} out of {\it four}  fundamental interactions of nature. These theories 
are characterized by the existence of first-class constrains on {\it them} in the terminology 
of Dirac's prescription for the classification scheme of constraints [1, 2].
This is the most modern definition of  any arbitrary $p$-form ($p = 1, 2, 3,...$)
gauge theory   in any arbitrary D-dimension of spacetime.
 There is a problem of canonical quantization in every gauge theory. To overcome this problem, one of the most intuitive mathematical  
approaches for the covariant canonical quantization of any gauge invariant field theory
 is the Becchi-Rouet-Stora-Tyutin (BRST) formalism [3-6]. This approach  not only serves as the foundation for renormalization programmes,  
but it has also been made in a huge variety of applications.
Gauge symmetry generalizes to BRST symmetry at the quantum level through the replacement of infinitesimal 
local gauge parameter by ghost and anti-ghost  variables/fields.  Therefore,  we have two types of quantum  symmetries 
(i.e., BRST and anti-BRST). The study of anti-BRST symmetry is important because it makes the theory complete and ghost free.
These symmetries are nilpotent of order two  and absolutely  anti-commuting which signify the  
 fermionic  nature  of symmetries and  linear independence of both of the symmetries, respectively.

 The BRST symmetry is characterized by two main features. The first is that it has a group theoretical basis: 
applying two different gauge transformations sequentially and then reversing their order does not produce the 
same outcome (unless the initial symmetry is Abelian), but the two distinct outcomes are connected by a group 
theoretical rule. The second significant property of the BRST transformations is nilpotency itself. It is
 obtained from the anticommuting properties of the ghost and anti-ghost fields via the Faddeev-Popov 
quantization method [7]. This indicates that if we perform the identical change again, we get zero. 
The Wess-Zumino consistency conditions, a key tool in the investigation of anomalies, are created when
 the two properties of BRST symmetry are combined. It should be observed that the second property is totally 
quantum, whereas the first property is classical. In other words, the BRST symmetry is quantum symmetry. 
The study of BRST formalism is very relevant because of the various applications of the (anti-)BRST symmetry 
transformations to many practical systems. The various issue connected with the anomalies have been discussed 
clearly [8-11], and the recent applications of BRST symmetry in M-theory [12, 13] and quantum gravity [14, 15]
 have also been discussed nicely within the framework of BRST.

The study of different quantum cosmological models is very interesting and challenging towards
 the development of a quantum theory of gravity where unification of general relativity and
 quantum mechanics is described [16, 17]. The description of homogeneous and isotropic 
spacetime symmetry was developed by Friedmann-Robertson-Walker (FRW) and the universe models
 associated with it  are  christened as FRW models [18-23]. The FRW models have played a crucial
 role in the development of the modern cosmology where most of works on quantum cosmology and 
dark energy in FRW spacetime are based on these models. Nevertheless, it is worthwhile to mention
 that most of the models of dark energy suffer from some problems related to cosmological constant 
(i.e., fine-tuning and coincidence problems, etc.). Therefore, it is important to do a more careful
 study of the basics of the cosmological FRW models in isotropic and homogeneous spacetime.
Some of the  BRST analysis has already been looked at for FRW models [24-28], we are highly motivated to carry 
out this research with some important and significant {\it novel} investigations for the cosmological FRW model 
within the realm of BRST formalism.

In this work, we will take BRST and also anti-BRST symmetries to analyze the fate of time parameterization invariance in a quantum theory for the 
cosmological FRW model. In fact, the fixing of gauge condition results in the breaking  of original gauge symmetry. Nevertheless, 
the extension of phase space via the introduction of additional variables allows us to recover a formulation with (anti-) BRST symmetries 
where transition amplitudes do not depend on the choice of gauge-fixing condition. Therefore, for the present investigation, the choice of the 
differential gauge-fixing condition  in extended phase space is appropriate from the theoretical point of view.
The extended phase space is formulated by considering the cosmological constant as a variable in the universe model.

The usual supervariable/superfield approach (USFA) to BRST formalism [29-33] uses the horizontality condition (HC) to
 derive off-shell nilpotent (anti-)BRST symmetry transformations for the gauge, ghost, and anti-ghost 
variables/fields when full super expansions of the supervariable/superfield [with two Grassmannian coordinates ($\theta, \bar\theta$)]
 are taken into account. In an interaction theory, the USFA does not explain how the (anti-)BRST symmetries for matter fields are derived. 
To overcome this problem, the idea of HC and gauge invariant restriction(s) (GIR) are combined together and this approach
 is generalized to obtain the (anti-)BRST symmetries for matter fields [34-36].
 The augmented version of the supervariable/superfield (AVSA) technique is the name given to this extended version of USFA.
 In light of the preceding discussions, we used the {\it newly} proposed (anti-)chiral supervariable/superfield  formalism (ACSA) 
 to derive the entire set of  nilpotent (anti-) BRST symmetries where (anti-)chiral super expansions of the supervariables/superfields 
 (with only {\it one} Grassmannian coordinate  in the super expansions of the supervariables) 
have been taken into account (see, e.g. [37-42] for details). 
In our present endeavor, for the first time, the analysis of the BRST invariant cosmological model (i.e., FRW model)
 is discussed in the extended phase space within the realm  of ACSA.
As far as, the application of the superfield method is concerned. The superfield approach nicely
applies to the description of consistent anomalies. It has been demonstrated that the superfield
 formalism not only makes it simple to recreate all the formulas relating to anomalies in any even dimension 
but also seems to be built specifically for them [8].

The different portions of this paper are organized as follows. In section 2, we discuss the constraints analysis  and 
gauge symmetry associated with the FRW model. The subject matter of section 3 deals with the construction of
gauge-fixing condition and (anti-) BRST analysis of the present model. In section 4 of this paper, we derive the Noether 
conserved (anti-)BRST conserved charges in extended phase space and demonstrate the nilpotency and absolute anti-commutativity
properties of these charges. Section 5 is fully devoted to the derivation of the nilpotent (anti-)BRST symmetries 
within the realm of ACSA to BRST formalism. Finally, in section 6, we summarize our
 key findings and draw conclusions about the novelty of our
work with some future investigations of cosmological models within the ambit of 
the supervariable/superfield approach to BRST formalism.\\

\vskip 0.5 cm

\section{Preliminaries: Constraints Analysis and Gauge Symmetries of the  FRW Model}

In this section, we will go over the preliminary of  the cosmological FRW model that describes a homogeneous and isotropic universe.
 We begin with the  FRW metric tensor  defined in spherical coordinates $(t, r, \vartheta, \phi)$ as follows;
\begin{eqnarray}
d s^2  =  N^2 (t) \, dt^2 + a^2(t) \,\Big(\frac {1} {1 - k r^2}\Big) \, d r^2 
 +  a^2(t) \,\Big( r^2 \,d\vartheta^2  + r^2 \,\sin^2\vartheta \, d\phi^2\Big),
\end{eqnarray}
where $N (t)$ denotes the lapse function and $a(t)$  is the scale factor of the universe  
that encrypts  the size at large scales.  The values of $k = 0, -1, +1$  correspond  to a space of
 zero curvature (i.e., flat universe), negative  curvature (i.e., open universe) and  positive curvature (i.e., closed universe) of universe,  respectively.

We now define the classical Lagrangian\footnote{This Lagrangian shows  a second-order  Lagrangian.
However, first-order Lagrangian   corresponding to the above Lagrangian  $L_s$ is  given, using the Legendre transformation as: 
${ L}_{f}  =   p_a\,\dot a + \frac {1}{2a}\, p_a^2\, N + \frac {1}{2}\,k\, N\,a,$
where $p_a ( = -\,  {a\,\dot a}/{N})$ denotes the canonical conjugate  momenta corresponding to the scale factor $a$.} 
of the FRW model represented in Arnowitt-Desert-Misner (ADM) variables as follows;
\begin{eqnarray}
{ L}_{s} =  -\frac{1}{2}\,\frac {a\,\dot{a}^2}{N}+\frac{1}{2}\,k\,N\,a,
\end{eqnarray}
where $\dot a$ denotes the time derivative of $a$ w.r.t. time $t$  (i.e., $\dot a = {da}/{dt}$). 
This Lagrangian holds the following Euler-Lagrange equations of motion (EL-EOMs):
\begin{eqnarray}
\dot a^2  + k\,N^2 = 0, \qquad a\, \ddot a + k\, N^2 + \frac {\dot a^2}{2} - \frac {a\,\dot a}{N} = 0.
\end{eqnarray}
 Now the canonically conjugate momenta corresponding to the lapse function $N$  turn out to be 
zero because  there is no time derivative of $N$  in  Lagrangian $L_s$, we get
\begin{eqnarray}
\Pi_{(N)} & \approx &0. 
\end{eqnarray}
This canonical conjugate momenta is weakly zero because the first-order time derivative is performed
which reflect the primary constraint of the theory. 
The canonical Hamiltonian of the this theory is given as: 
\begin{eqnarray}
H = p_a\, \dot a - L_f = -\, \frac {N\, p_a^2}{2\,a} - \frac {1}{2}\, k\,N\,a.   
\end{eqnarray}
Using the time conservation of the primary constraint,  we calculate the  secondary-constraint of the theory as follows:
\begin{equation}
\frac {d\Pi_{(N)}}{dt} = \frac{p_a^2}{2a}+\frac{k}{2}a  \approx  0.
\end{equation}
 Both the constraints are first-class as they commute with each other, which confirms that the cosmological 
FRW model is endowed with gauge symmetry. The gauge symmetry of the variables present in Lagrangian $L_s$ 
associated with this model is given by   
\begin{equation}
\delta_g N = - N\,\dot\eta - \dot N\,\eta,\qquad\qquad  \delta_g a = - \dot a\,\eta,
\end{equation}
where  $\eta(t)$ is an infinitesimal   gauge transformation parameter.
Under the above infinitesimal gauge transformations  Lagrangian $L_s$ transforms to a  total time derivative as: 
\begin{eqnarray}
\delta_g L_s  =  \frac {d}{dt}\Big[-\, \eta\,L_s\Big]. 
\end{eqnarray}
Hence the Lagrangian (2) remains invariant  under the gauge transformations (7) which implies
the action integral of this model remains invariant.\\

\section {Lagrangian Formulation: Gauge-Fixing Condition}

\vskip 0.2 cm

Gauge fixing is a mathematical technique for dealing with extra degrees of freedom for field variables in gauge theory physics.
For the canonical quantization of any gauge  theory and  to remove the redundancy in gauge degrees of freedom, 
we use the traditional approach of imposing gauge-fixing conditions.
The following are the basic prerequisites for a  gauge-fixing condition of a gauge theory:
 (i)  it must entirely fix the gauge, i.e., there must be no residual gauge freedom, and 
(ii) on the application of  the  transformations,  it  must be able to bring any configuration given by $N$
 and $a$ into one satisfying the gauge condition.  Keeping the above parameters in mind, we select the following gauge condition
for the present FRW model in extended phase space;
\begin{equation}
\dot{N} =\frac{d}{dt}F(a),
\end{equation}
where $F(a)$ is an arbitrary function of $a$.
The above  gauge condition has been examined in order to provide 
a well-defined formulation in extended phase space. 
Since simply extending the phase  space by including  gauged degrees of freedom is insufficient. 
Therefore, Lagrangian should additionally include missing velocities terms, too. It is possible to do so using 
differential gauge condition, which actually extend  the phase space. The gauge condition in differential   form  is
\begin{eqnarray}
\dot N = \frac {d F}{d a}\,\dot a.
\end{eqnarray}
This gauge condition can be used in quantum theory by including the following 
gauge-fixing term in the invariant Lagrangian (2);
\begin{eqnarray}
{L}_{gf}  =   \lambda\left( \dot{N} - \frac {d F}{d a}\,\dot a\right), 
\end{eqnarray}
where $\lambda$ is an auxiliary variable  (i.e., Lagrange multiplier) used to linearize the gauge-fixing term.
Now the Faddeev-Popov  ghost terms of the FRW model corresponding to the above gauge-fixing term is given as  
\begin{eqnarray}
{L}_{gh}  =   \dot{\bar{C}}\left( \dot{N} -  \frac {d F}{d a}\,\dot a\right)\,C  + \dot{\bar{C}}\,N\,\dot{C},\label{gh}
\end{eqnarray}
where ($\bar C, \, C$) are  anti-ghost and ghost variables  having ghost number $(-1, \, +1)$, respectively. 
The above ghost terms are used for the consistency of the theory (i.e., to get rid of the unphysical degrees of freedom). 
Now, the complete extended Lagrangian for the theory ${L}_{ext}$ (i.e., ${L}_{s} + {L}_{gf} + {L}_{gh}$) reads:
\begin{eqnarray}
{L}_{ext}  & = &   -\, \frac{1}{2}\frac{a\dot{a}^2}{N} + \frac{k}{2}Na +  \lambda\left(\dot{N} - \frac{dF}{da}\dot{a} \right) 
 +  \dot{\bar{C}}\left( \dot{N} - \frac{dF}{da}\dot{a} \right)\, C + \dot{\bar{C}}\, N\, \dot{C}.
\end{eqnarray}
The BRST and anti-BRST  symmetry transformations corresponding to the above extended Lagrangian $L_{ext}$ are given as,
\begin{eqnarray}
s_b N  =  -\, (\dot{N}  C + N\dot C),\quad\; s_b a  =   -\, \dot{a}\, C,\quad \; s_b  C  =  -\,C\dot {C}, \quad s_b\bar C  
= -\, \lambda,\quad \; s_b \lambda  =  0,\nonumber\\
s_{ab} N  =  -\,(\dot{N}  \bar C + N \dot {\bar  C}),  \quad s_{ab} a  =  -\, \dot{a}\, \bar C,\quad \; s_{ab}  \bar C 
 = -\,\bar C\,\dot {\bar C},\quad s_{ab}\bar C  = \lambda,\quad s_{ab} \lambda  =  0.
\end{eqnarray}
It is straightforward to check that the above BRST and anti-BRST symmetry 
transformations  are supersymmetric  type (i.e., bosonic variables  transform to the fermionic variables and vice-versa),  
 nilpotent of order two (i.e., $s_b ^2 = 0, \; s_{ab} ^2  = 0)$
and absolutely anti-commuting with each other (i.e., $s_b\,s_{ab} + s_{ab}\,s_b = 0$).  
The above  extended Lagrangian  ${L}_{ext}$ is invariant under the (anti-)BRST symmetries  up to total time derivative.

The combination of gauge-fixing and ghost terms for both gauges are
BRST and anti-BRST exact, therefore, we have the following  
\begin{eqnarray}
{L}_{gf} +{L}_{gh} & = &  -s_b\left[\bar C\left( \dot{N} - \frac {d F}{d a}\,\dot a\right)\right]
 \;\;  \equiv \;\;  - s_{ab} \left[C\left( \dot{N} - \frac {d F}{d a}\,\dot a\right)\right],
\end{eqnarray}
which is one of the possible way to derive the gauge-fixing and ghost terms
using the above (anti-)BRST symmetry transformations. 
Under the above fermionic  symmetry transformations (14), the extended Lagrangian transform  as  the total time derivative:
\begin{eqnarray}
s_b L_{ext}  = -\, \frac {d}{dt}\Big[(\lambda + \dot{\bar C}\, C)\, 
\Big\{\Big(\dot N - \frac {d F}{d a}\,\dot a\Big)\, C + N\, \dot C \Big\} \Big], \nonumber\\
s_{ab} L_{ext}  =  \; \;\frac {d}{dt}\Big[(\lambda +   \bar C\, \dot{ C})\, 
\Big\{\Big(\dot N - \frac {d F}{d a}\,\dot a\Big)\, \bar C -  N\, \dot {\bar C} \Big\} \Big]. 
\end{eqnarray}
Hence the action integral (i.e. $S = \int \, d\,t\;L_{ext})$ corresponding to the  FRW model remains invariant and 
extended Lagrangian (13) is (anti-)BRST invariant Lagrangian. 
\\

\section {(Anti-)BRST Charges of the FRW Model}

\vskip 0.2 cm 

The Noether theorem states that if any Lagrangian or its corresponding action stays invariant
under any continuous symmetry transformation, there are conserved current and charges corresponding 
to that given continuous symmetry. Therefore, for the present FRW model, there exist BRST $(Q_b)$ and 
anti-BRST $(Q_{ab})$ charges given as:  
\begin{eqnarray}
 Q_b & = &  \frac{N\, C}{2a}\left[(\lambda + \dot {\bar C}\, C)\,F_a - \frac {a\,\dot a}{N}\right]^{2}
 +   \frac{k}{2}\, N\, C\,a +  \lambda\, \dot N\, C  - N\, \dot {\bar C}\, \dot C\, C, \nonumber\\
 Q_{ab} & = & \frac{N\, \bar C}{2a}\left[(\lambda +  \bar C\, \dot{ C})\,F_a + \frac {a\,\dot a}{N}\right]^{2} 
 +   \frac{k}{2}\, N\, \bar C\,a + \, \lambda\, \dot N\, \dot {\bar C} - N\, \dot {\bar C}\, \bar C\, \dot C, 
\end{eqnarray}
where $F_a$ is derivative of function $F$ w.r.t. $a$ (i.e.,  $F_a = {dF}/{da}$).  These charges act as the generators for the off-shell nilpotent (anti-)BRST symmetry transformations.
The conservation law ($\partial_t Q_b = 0,\; \partial_t Q_{ab} = 0$) of these 
charges can be proven by using the following EL-EoMs evaluated form the extended 
Lagrangian (13), namely;
\begin{eqnarray}
&& \dot \lambda + \ddot{\bar C}\, C - \frac {a\, {\dot a}^2}{2\, N^2} - \frac {k\,a}{2} = 0,\nonumber\\
 &&\dot N -  F_a\, \dot a = 0, \quad \ddot {\bar C}\, N + \dot {\bar C}\,F_a\, \dot a = 0,\nonumber\\
&&  2\, \ddot N\, C + 2\, \dot N\, \dot C - \dot F_a\, \dot a\, C - F_a\, \ddot a\, C- F_a\, \dot a\, \dot C = 0,\nonumber\\
  &&\frac {a\,\dot a}{2N^2}  -\, \frac {a\,\ddot a}{N} - \frac {k\,N}{2}
\lambda \, (\ddot N -  \dot F_a) + \dot \lambda\, (N -  F_a)+ \ddot N\, \dot {\bar C}\, C  \nonumber\\
  && \dot N\, \ddot {\bar C}\, C  + \dot N \dot {\bar C} \dot C - \dot F_a\, \dot {\bar C} C 
- F_a\, \ddot {\bar C}\, C - F_a\, \dot {\bar C}\, \dot C = 0.~~ 
\end{eqnarray}
It is straightforward to check that the above (anti-) BRST conserved charges [${Q}_{(a)b}$] are nilpotent 
 of order two (i.e., ${Q}_b^2 = {Q}_{ab}^2 = 0)$ and they follow absolute anti-commutativity property 
(i.e., $Q_b \, {Q}_{ab} + {Q}_{ab}\,{Q}_b = 0$) in the extended phase space. 
These two features are captured by the definition of generator, which is as follows:
\begin{eqnarray}
&&s_b {Q}_b \;\;\,\,  =\; -\,i\,\,{\{{Q}_b,{ Q}_b}\} = 0 \;\;\,\; \Longrightarrow \;\; \, {Q}_b^2 = 0,\nonumber\\
&&s_{ab} \,{Q}_{ab} =\; -\,i\,\{{ Q}_{ab},{Q}_{ab}\} = 0 \,\;\Longrightarrow \,\;\;  {Q}_{ab}^2 = 0,\nonumber\\
&&s_{ab}{Q}_{b} \;\; =\; -\,i\,{\{{Q}_{b},{Q}_{ab}}\} = 0 \;\;\; \Longrightarrow \;\; {Q}_{b}\,{Q}_{ab} + {Q}_{ab}\,{Q}_{b} = 0, \nonumber\\ 
&& s_b {Q}_{ab}  \;\; =\; -\,i\,{\{{Q}_{ab},{Q}_b}\} = 0 \;\;\; \Longrightarrow \;\; {Q}_{ab}\,{Q}_{b} + {Q}_{b}\,{Q}_{ab} = 0. 
\end{eqnarray}
The above properties describe  the nature and identity of the conserved (anti-)BRST charges of the cosmological FRW model.
  The nilpotency property of these conserved  charges signifies  the fermionic nature  whereas
absolute anti-commutativity property, physically, shows that they are linearly independent 
with each other.    
\\

\section {(Anti-)BRST Symmetry Transformations: ACSA}

\vskip 0.2 cm

In this section, we provide  explicit and   step-by-step derivation of the off-shell nilpotent  BRST and anti-BRST symmetry transformations 
for the cosmological FRW model in the extended phase space. Toward this aim in mind, first of all, 
we generalize our basic variables [i.e., $N(t),\; a(t),\; C(t), \; \bar C(t)$] and auxiliary variable [i.e., $\lambda (t)$] 
 of Lagrangian [(cf. Eq. (13)]   onto (1, 1)-dimensional (anti-)chiral    super-subspace of the most  general 
(1, 2)-dimensional superspace by addition of one of the
 Grassmannian coordinate  (i.e. $\theta, \, \bar\theta$) into the given time dimension coordinate. 
We assume that $s_b \bar C = -\lambda $ and $s_{ab} C = \lambda $ where $\lambda$ is the Nakanishi-Lautrup type
auxiliary variables used to linearize the differential gauge-fixing term. These BRST and anti-BRST symmetry transformations (i.e. $s_b \bar C = -\lambda,\; s_{ab} C = \lambda$) are the
standard assumptions in the realm of BRST formalism.

First of all, we focus on the derivation of BRST symmetry transformations of the FRW model. 
For this, we generalize all the basic and auxiliary  variables  present in the Lagrangian $L_{ext}$ onto the 
anti-chiral supervariable as 
\begin{eqnarray}
&& a(t) \;\; \longrightarrow \; \;A (t, \bar\theta) \;\; = \; a(t) + \bar\theta\, b_1 (t), \nonumber\\
&& N(t) \; \longrightarrow \;\; {\cal N} (t, \bar\theta) \;=   \;  N(t) + \bar\theta\, b_2 (t), \nonumber\\
&& C(t) \; \longrightarrow \;\; F (t, \bar\theta) \;\; = \;  C(t) + \bar\theta\, f_1 (t), \nonumber\\
&& \lambda (t) \;\; \longrightarrow \;\; \Lambda (t, \bar\theta) \;\; = \;  \lambda (t) + \bar\theta\, b_3 (t), 
\end{eqnarray}  
where $b_1 (t),\; b_2 (t),\; b_3 (t)$ are the bosonic secondary variables and $f_1 (t)$ is the 
fermionic secondary variable which we  have to be determined using the standard technique of anti-chiral supervariable approach.
The fermionic and bosonic natures of these secondary variables are ensured by the fermionic nature of Grassmannian variable $\bar\theta$.

We employ the highly important and intriguing  BRST invariant quantities, which 
are the combinations of the basic and auxiliary variables of the Lagrangian (13),
to derive the aforementioned secondary variables, namely;
\begin{eqnarray}
&& s_b \lambda = 0, \quad s_b ( C\, \dot C) = 0,\quad s_b (C\, \dot a) = 0,\quad s_b (N\,\dot C + \dot N\, C) = 0.
\end{eqnarray} 
According to the basic premise of the ACSA to BRST formalism,
the above set of BRST invariant restrictions must be independent of the Grassmannian  coordinate  $\bar\theta$
 when these invariant quantities are generalized with the coordinate  $\bar\theta$,  we have 
\begin{eqnarray}
&& \Lambda (t, \bar\theta) = \lambda (t), \quad F (t, \bar\theta)\, \dot  F(t, \bar\theta) = C (t)\, \dot  C (t), \quad F (t, \bar\theta) \, \dot {A} (t, \bar\theta) =  C (t)\, \dot a (t),\nonumber\\
 && {\cal N} (t, \bar\theta)\, \dot F (t, \bar\theta) + \dot {\cal N} (t, \bar\theta)\, F (t, \bar\theta)\nonumber = N (t)\,\dot C (t) + \dot N (t)\, C (t).
\end{eqnarray} 
Now using the generalization of non-trivial BRST invariant restriction $s_b \lambda = 0$, we have 
\begin{eqnarray}
s_b \lambda = 0 \;\; \Longleftrightarrow \;\;  \Lambda (t, \bar\theta) = \lambda (t) \;\; \Longrightarrow \;\;  b_3 (t) = 0.
\end{eqnarray}
To derive the value of other secondary variables, first of all, we use the trivial BRST invariant
 quantity  $s_b ( C\, \dot C) = 0$, we get the following generalization:  
\begin{eqnarray}
s_b (C\, \dot  C) = 0 \;\; & \Longleftrightarrow & \;\; F (t, \bar\theta)\, \dot  F(t, \bar\theta) =  C (t)\, \dot C (t)\nonumber\\
 & \Longrightarrow & \dot C (t)\, f (t)  + C(t)\, \dot f_1 (t) = 0 \nonumber\\
 & \Longrightarrow & f_1 (t) = -\, C (t)\, \dot  C(t). 
\end{eqnarray}
Substituting the above  value of secondary variable $f_1 (t)$ into the anti-chiral super expansion of the supervariable (20), we have 
\begin{eqnarray}
C(t) \; \longrightarrow \; F^{(b)} (t, \bar\theta) = C(t) + \bar\theta\,[-\, C (t)\, \dot C(t)]. 
\end{eqnarray} 
Now using the generalizations of the trivial BRST invariant restrictions  $s_b (C\, \dot a) = 0$ and $ s_b (N\,\dot C + \dot N\, C) = 0$ which 
give the following expression 
\begin{eqnarray}
&& s_b (C\, \dot a) = 0  \;  \Longleftrightarrow \;  F^{(b)} (t, \bar\theta) \, \dot {A} (t, \bar\theta) =  C (t)\, \dot a (t) \;\;\; \Longrightarrow \;  b_1 (t) = -\, \dot a(t)\, C(t), \nonumber\\
&& s_b (N\,\dot C + \dot N\, C) = 0    \Longleftrightarrow     {\cal N} (t, \bar\theta)\, \dot F^{(b)} (t, \bar\theta) + \dot {\cal N} (t, \bar\theta)\, F^{(b)} (t, \bar\theta)  =  N (t)\,\dot C (t) + \dot N (t)\, C (t)\nonumber\\
&& \Longrightarrow  b_2 (t) = -\,[N(t)\, \dot C (t) + \dot N(t)\, C(t)],  
\end{eqnarray}
where superscript $(b)$ on the supervariable $F (t, \bar\theta)$ denotes the supervariable has been obtained after the 
application of the BRST invariant quantity.  
Thus, finally, we have the following  expressions for the anti-chiral super expansions of the supervariable
\begin{eqnarray}
 a(t) \; \longrightarrow \; A^{(b)} (t, \bar\theta) \; & = & a(t) + \bar\theta\,[-\, \dot a(t)\, C(t)] =  a(t) + \bar\theta \,(s_b \,a), \nonumber\\
 N(t)  \longrightarrow \; {\cal N}^{(b)}   (t, \bar\theta) & = & N(t) + \bar\theta\, [-\,(N(t)\, \dot C (t) +  \dot N(t)\, C(t))]\nonumber\\ 
& = & N(t) + \bar\theta\,(s_b N), \nonumber\\
 C(t) \, \longrightarrow \; F^{(b)}   (t, \bar\theta) \; & = & C(t) + \bar\theta\, [-\,C(t)\, \dot  C(t)] =  C(t) + \bar\theta\, (s_b C), \nonumber\\
\lambda (t)  \;\, \longrightarrow \; \Lambda ^{(b)}  (t, \bar\theta) \; & = & \lambda (t) + \bar\theta\, [0] =  \lambda (t) + \bar\theta \,(s_b \lambda), 
\end{eqnarray} 
where superscript $(b)$, once again,   on all the supervariables denote the supervariables have been obtained after the 
use of BRST invariant restrictions (21). Here we found that the coefficients of the $\bar\theta$ are nothing but the 
BRST symmetries  of the basic and auxiliary variables  of the theory. 
Therefore, we have a concluding remark that $s_b$ is connected with the translational generator  $\partial_{\bar\theta}$ along the 
$\bar\theta$-direction as: $s_b\longleftrightarrow \partial_{\bar\theta}$ (see, e.g. [29-33]).

Now we are in position to derive the anti-BRST symmetry transformations for various variables  of the Lagrangian (13) 
within the ambit of ACSA to BRST formalism. Towards this aim in mind, first of all, we generalize our basic and auxiliary 
variables by adding one Grassmannian variable $\theta$ into the ordinary variables, we get chiral super expansions of the supervariables, namely;
\begin{eqnarray}
&& a(t) \; \,\;\longrightarrow \; A (t, \theta) \, = a(t) + \theta\, \bar b_1 (t), \nonumber\\
&& N(t) \; \longrightarrow \; {\cal N} (t, \theta) = N(t) + \theta\, \bar b_2 (t), \nonumber\\
&& \bar C(t) \; \; \longrightarrow \; \bar F (t, \theta) \,= \bar C(t) + \theta\, \bar f_1 (t), \nonumber\\
&& \lambda (t) \; \; \, \longrightarrow \; \Lambda (t, \theta) \;= \lambda (t) + \theta\, \bar b_3 (t), 
\end{eqnarray}  
where $\bar b_1 (t),\; \bar b_2 (t),\; \bar b_3 (t)$ are the bosonic secondary variables and $\bar f_1 (t)$ is the 
fermionic secondary variable which we is  determined by using the standard technique of chiral supervariable approach.
For this, once again, we exploit very important and interesting anti-BRST invariant quantities given below, 
\begin{eqnarray}
&& s_{ab} \lambda = 0, \quad s_{ab} ({\bar C}\, \dot {\bar C}) = 0,\quad s_{ab} (\bar C\, \dot a) = 0,  \quad 
 s_{ab} (N\,\dot {\bar C} + \dot N\, {\bar C}) = 0.
\end{eqnarray} 
According to the basic premise of ACSA to BRST formalism any anti-BRST invariant quantity must be independent of the 
Grassmannian variable $\theta$. Therefore the above anti-BRST invariant restrictions (Eq. (29)) generalize in the 
following fashion: 
\begin{eqnarray}
&& \Lambda (t, \theta) = \lambda (t), \quad {\bar F} (t, \theta)\, \dot  {\bar F} (t, \theta) = {\bar C} (t)\, \dot  {\bar C} (t),\nonumber\\
&& {\bar F} (t, \theta) \, \dot {A} (t, \theta) =  {\bar C} (t)\, \dot a (t),\quad {\cal N} (t, \theta)\, \dot {\bar F} (t, \theta)\nonumber\\
&& + \dot {\cal N} (t, \theta)\, {\bar F} (t, \theta) = N (t)\,\dot {\bar C} (t) + \dot N (t)\, {\bar C} (t).
\end{eqnarray}
The above generalized anti-BRST invariant quantities lead to the derivation of the following
expressions of the secondary variables: 
\begin{eqnarray}
&& \bar b_1 (t) = -\, \dot a (t) \, \bar C (t), \;\; \bar b_2 (t)  = -[\dot N (t) \, \bar C (t) + \dot N (t) \, \bar C (t)],\nonumber\\ 
&& \bar f_1 (t)  =  -\,{\bar C}(t) \, \dot {\bar C} (t), \;\;  \bar b_3 (t)  = 0.  
\end{eqnarray}
After substituting the above value of secondary variables into the chiral  super expansions
 (28) of the  chiral supervariables, we get the following 
\begin{eqnarray}
a(t) \;\; \longrightarrow \;\;  A^{(ab)} (t, \theta) \;  &=& \; a(t) + \theta\,[-\, \dot a(t)\, \bar C(t)] =  a(t) + \theta \,(s_{ab} a), \nonumber\\
N(t) \; \longrightarrow \;\;  {\cal N}^{(ab)}   (t, \theta) \; &=& \; N(t) + \theta\, [-\,\{\dot N(t)\, \dot {\bar C} (t) +  \dot N(t)\, {\bar C}(t)\}]\nonumber\\ 
& = & N(t) + \theta\,(s_{ab} N), \nonumber\\
\bar C(t) \; \longrightarrow \; \; \bar F^{(ab)}   (t,\theta) \;  &=& \; \bar C(t) + \theta\, [-\,{\bar C}(t)\, \dot {\bar C}(t)] =  \bar C(t) + \theta\, (s_{ab} \bar C), \nonumber\\
 \lambda (t) \;\; \longrightarrow \; \;\Lambda ^{(ab)}  (t, \theta) \;  &=& \; \lambda (t) + \theta\, [0] =  \lambda (t) + \theta \,(s_{ab} \lambda), 
\end{eqnarray} 
where superscript $(ab)$ on the chiral supervariables denote the supervariables that have derived after the use of 
anti-BRST invariant restrictions (29). Here, the coefficients of  Grassmannian variable $\theta$ denote 
the anti-BRST symmetry transformations for the various basic and auxiliary variables of the cosmological FRW model. 
We end this section with a concluding remark that $s_{ab}$ is deeply linked with  the translational generator 
$\partial_\theta$ along the $\theta$-direction of the Grassmannian variable with a mapping: $s_{ab} \longleftrightarrow \partial_\theta$ (see, e.g. [29-33] for detail). 
Therefore, the BRST and anti-BRST symmetries acting on the variables are geometrically connected with the translations of supervariables
after the application of translational generators $(\partial_{\bar\theta}, \partial_\theta)$ along the 
Grassmannian coordinates ($\bar\theta, \theta$), respectively.\\

\section {Conclusion}

he cosmological FRW models are well-known representations of a homogeneous and isotropic universe  in the extended phase space 
where we assume zero cosmological constant, therefore, the fundamental interaction of gravity is the only force at work.
These models are of utmost significance in the contemporary cosmology because they are  relevant  
in the  behavior and  evolution of universe. 
At present, most of the investigations connected with dark energy  are described by these cosmological models in the FRW universe.

In this research, we looked at FRW model that described a closed, flat, and open universe.
We found that cosmological FRW model endowed with the first-class constraints in the language 
of Dirac's prescription of classification scheme of constraints. We quantize this model
 using the BRST analysis with differential gauge 
condition  in the extended phase space. We obtained the BRST and anti-BRST symmetry 
transformations by the replacement of the  local gauge parameter through the ghost and anti-ghost variables,  respectively. 
We develop, for the first time,  the anti-BRST symmetry transformation for the cosmological FRW model which is  
 useful in explaining the full geometrical description of both the symmetries  in the realm
 of supervariable/superfield approach.

We derived the nilpotent BRST and anti-BRST conserved charges associated with the FRW model
 using the  Noether theorem where global (anti-)BRST invariance
 of the effective action in the extended phase space play important role. We also demonstrated nilpotency and absolute anti-commutativity 
properties of these charges within the framework of BRST formalism. 
To prove the sanctity of the (anti-)BRST symmetry transformations associated with the present FRW model, 
we have derived the complete set of symmetries within the ambit of (anti-)chiral supervariable approach 
(ACSA) to BRST formalism where only one Grassmannian variable has been taken into account
whereas in usual supervariable/superfield  approach [29-33] full super expansion of the supervariable/superfield
with two Grassmannian coordinates have been taken into account.  
For the present endevour,  we discovered that BRST symmetry $(s_b)$ is linked to the translational generator $(\partial_{\bar\theta})$
 along the $\bar\theta$-direction, whereas anti-BRST symmetry $(s_{ab})$ is linked to the translational 
generator $(\partial_\theta)$ along the $\theta$-direction of the Grassmannian variables.

Our future investigations will focus on expanding the scope of this study to more complex cosmological
 models to see how well the BRST framework can achieve gauge and diffeomorphism invariance at the quantum level.
It would be fascinating to discuss   the (anti-)BRST symmetry transformations associated with 
a more general form of the cosmological FRW model where 1D reparameterization (i.e., diffeomorphism) 
symmetry plays an important role.  The combination of the ACSA and the modified Bonora-Tonin 
supervariable approach (MBTSA) would be very interesting to discuss and derive the (anti-)BRST 
symmetry transformations and many more properties associated with for FRW model. It would be very 
significant step toward the basic understanding and formation of a complete quantum theory of the modern cosmology.
In future endeavors, we would also like  to work on the physical implications  of our results (e.g. anomalies, M-theory
 and quantum gravity) for the various practical systems/models within the framework of BRST  
and supervariable/superfield  formalism.\\

 \vskip 0.5cm

\noindent
{\bf\large Data Availability}\vskip 0.3cm

\noindent
No data were used to support this study.\\ 

\vskip 0.3cm 

\noindent
{\bf\large Conflicts of Interest}\vskip 0.3cm

\noindent
The authors declare that there is no conflicts of interest.\\ \vskip 0.2cm

\noindent
{\large\bf Acknowledgments}\\

\noindent
The current investigation  has been carried out under the  DST-INSPIRE fellowship (Government of India)  awarded to the author.
The author is extremely grateful to the above fellowship program for its financial support.  
 The author is also thankful  to A. K. Rao for a careful reading of this manuscript.
Fruitful and enlightening comments ans suggestions by our esteemed Reviewer is thankfully acknowledged.

\end{document}